\documentclass[a4paper]{article} 
\pdfoutput=1

\usepackage{authblk} % for affilations 

\usepackage[pdftex]{graphicx} %graphics in pdfLaTeX; include pictures

\usepackage[thickqspace,squaren]{SIunits} 
  \usepackage{amsmath} 

\title{Quench Dynamics of Two Coupled Zig-Zag Ion Chains}
\author[1]{Andrea Klumpp\footnote{andrea.klumpp@physnet.uni-hamburg.de}}
\affil[1]{Zentrum f\"{u}r Optische Quantentechnologien, Universit\"{a}t Hamburg, Luruper Chaussee 149, 22761 Hamburg, Germany}

\author[2]{Benno Liebchen}
\affil[2]{SUPA, School of Physics and Astronomy, University of Edinburgh, Edinburgh EH9 3FD, United Kingdom}

\author[1,3]{Peter Schmelcher}
\affil[3]{The Hamburg Centre for Ultrafast Imaging, Luruper Chaussee 149, 22761 Hamburg, Germany}

\begin{document}

\maketitle

\begin{abstract}
We explore the non-equilibrium dynamics of two coupled zig-zag chains of trapped ions in a double well potential. 
Following a quench of the potential barrier between both wells, the induced coupling between both chains due to the
long-range interaction of the ions leads to their complete melting. 
 The resulting dynamics is however not exclusively
irregular but leads to phases of motion during which various ordered structures appear with ions arranged in 
arcs, lines and crosses. We quantify the emerging order by introducing a suitable measure 
and complement our analysis of the ion dynamics using a normal mode analysis showing a decisive population transfer between only a few distinguished modes.
\end{abstract}

% \begin{keyword}
% ion trapping \sep nonlinear dynamics \sep chaos
% \end{keyword}

\section{Introduction}
Cooled ions in traps form a clean and highly versatile setup for exploring structure formation with long-range interacting particles, both in equilibrium and non-equilibrium. In equilibrium, possible structures include small and large ion crystals \cite{Died1987,Wine1987,Drew1998}
possessing various internal ordering such as concentric rings (2D), shells (3D) \cite{Boll1994,Dub1999, Boni2008} and string-of-disks 
configurations \cite{Drew2003}, and even two-component Coulomb bicrystals \cite{Drew2001}. At the crossover from one to higher dimensions trapped ions can also form zig-zag configurations, a structure that attracts
particular attention in the recent literature \cite{Mori20041,Mori20042,Balt20121}. Specifically, varying the geometry of an anisotropic harmonic trap allows for a second-order phase transition from a linear to a zig-zag structure, which can be either ideal or involve topological defects \cite{Dub1993,Schiff1993,Miel2012}.
Out of equilibrium, the recent literature predicts an equally rich variety of possible ionic structures including 
spatiotemporal patterns in laser-driven microtraps \cite{Lee2011} and periodic lattices \cite{Liebchen2014,Liebchen2015},
but also interaction induced current reversals of the transport direction \cite{Liebchen2012} based on structure formation in the phase space.
Much of this research on structure formation with trapped ions roots in the admirable advancements of the controllability of ions in recent years. 
This ranges from the quickly progressing miniaturization of ion traps and 
lab on chip technologies \cite{Hugh2011, Wilp2012} via the advent of optical trapping techniques \cite{Scha2010} to the discovery of multi-segmented Paul and Penning traps \cite{Schu2008,Tana2014}.
The latter example in particular allows for more and more complex but still controllable arrangements of long-range interacting particles as required e.g. for quantum information processing \cite{Gui2015,Kauf2014,Schu2006,Blatt2008}.
The above advancements allow and evoke a new type of question: How do individual ionic structures respond if we couple them to each other? 
Consider for example a segmented ion trap with two wells, both loaded with an individual zig-zag 
configuration, separated from each other by a potential barrier between the wells.
Let us now quench the barrier to a lower value, which increases the coupling between the individual chains:
Are the only two (expected) alternatives for the dynamics of the zig-zag chains that they either
deform only slightly and respond with small oscillations to the increased coupling 
or that we observe their complete melting resulting in irregular oscillations 
of all ions? This is precisely the problem we want to investigate in the present work.
To explore the above problem we develop a minimal model based on a two dimensional double well potential that allows for zig-zag configurations in both wells whose geometries resemble
the well-known zig-zag states in anisotropic single well traps. 
Remarkably, following the quench of the barrier we observe that the melting process does not simply lead
to irregular oscillations but to a complex non-equilibrium dynamics
constituted of different phases of motion. Phases of irregular oscillatory motion are interrupted by motional phases
which exhibit transient ordered
configurations. Although nonlinear dynamics governs the motion of the coupled
ion chains, 
 we employ a normal mode analysis showing that the population of the corresponding linear eigenvectors is not arbitrarily distributed
over the whole band of modes as one would expect for e.g. a chaotic system.
Instead, during most phases of the time evolution only a few eigenvectors are strongly populated
and nonlinear effects show up in form of a decisive and quite sudden transfer of energy among the different modes.
Our work is organized as follows. Section II explains our setup and the preparation of the ground state configuration.
Section III provides our main results followed by a normal mode and population analysis of the dynamics. We summarize our findings
and their interpretation in section IV.

\begin{figure}[h!]
\begin{center}
 \includegraphics[width=0.8 \linewidth]{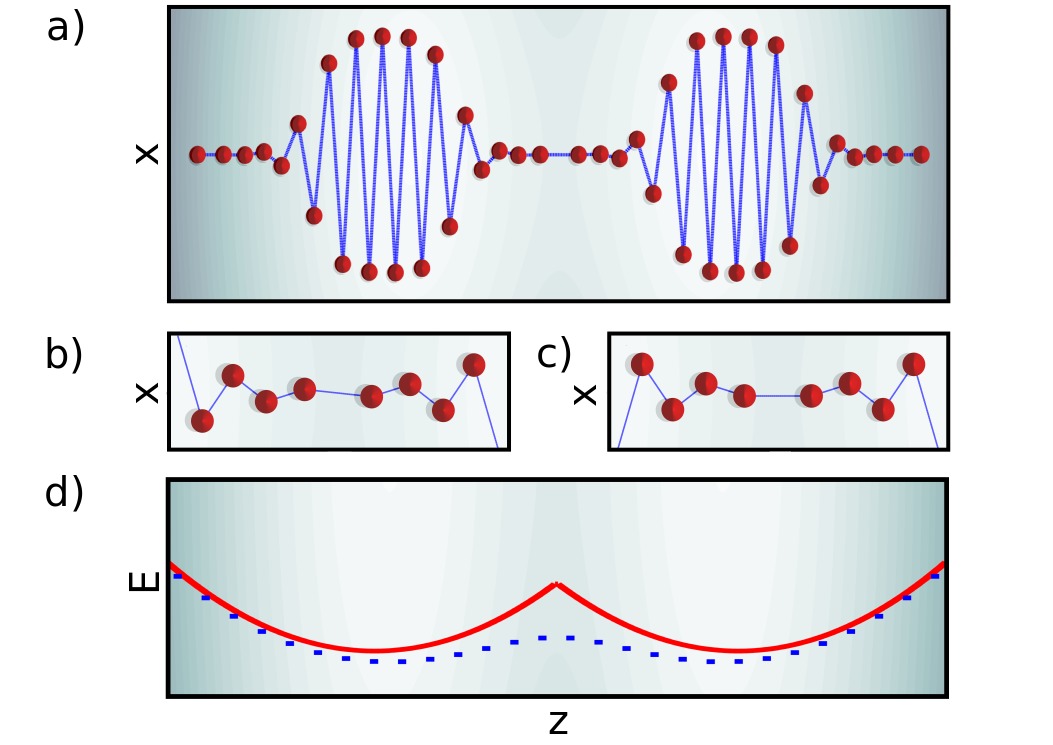}
\caption{{\small (a) Cartoon of the double zig-zag equilibrium configuration in the double well potential used as the initial configuration.
(b,c) Magnifications of the part of the ionic configuration which links between the two chains, i.e. in
the barrier region. It highlights the difference between the ground state configuration (b) and the energetically next higher equilibrium configuration (c) 
which we call the mirror configuration. (d) The red (solid) line shows the double well potential before the quench and the blue (dashed)
line after the quench. A lowered barrier enhances the coupling between the two ion chains (arbitrary units throughout).}}
\label{figa}
\end{center}
\end{figure}

\section{Setup, Hamiltonian and ground state configuration}
We consider $N$ ions, described as classical point particles with mass $m$ and charge $Q$, confined in radial direction ($x,y$) to a 
linear quadrupole Paul trap and to a double well potential (segmented trap) in $z-$direction. 
\footnote{Note that formally, the Laplace equation does not allow to combine an ideal linear Paul trap with a double well potential in axial direction. However, 
experimental traps are not ideal and reference \cite{Rust2014} could indeed realize the combination of an 
approximately linear Paul trap and a Mexican hat like potential which justifies to consider a combination of a double well potential with a linear Paul trap.}
\begin {equation} 
\Phi(x,y,t) =\frac{U_{\rm dc}}{2}(cx^2+cy^2)+\frac{U_{rf}}{2}\cos{(\omega_{\rm rf} t)}(cx^2-cy^2). 
\end {equation}
with $U_{\rm dc}$ and $U_{\rm rf}$ being the applied constant and the rf-voltage; $\omega_{\rm rf}$ is the (radio)frequency, c is a geometrical parameter of the trap.
The geometrical parameter is for both directions x and y equal in radially symmetric traps but in planar traps \cite{Mad2004,Tana2014} the geometrical parameter for both directions can strongly differ. 
The ion dynamics in the radio-frequency trap is composed of the so-called micro motion, and a comparatively slow averaged motion taking place in an effective harmonic potential \cite{Ger1992}
$ V(x,y)=\frac{m}{2}(\omega_x^2 x^2+\omega_y^2 y^2)$.
Here, $\omega_{x}=\frac{\omega_{rf}}{2}\sqrt{a-q^2/2}$ and $\omega_{y}=\frac{\omega_{rf}}{2}\sqrt{a+q^2/2}$ are the effective trapping frequencies with 
$a=\frac {4QU_{dc}}{m \omega^2_{rf}}c$ and $q=\frac{2QU_{rf}}{m\omega_{rf}^2}c$ being dimensionless parameters. For the confinement in $z$-direction we assume the following phenomenological double well potential \cite{Stre2006}, with wells 
centered at $\approx\pm z_0$ and separated from each other by a barrier of height $\sim 1/C$ (see Fig.~\ref{figa}). 
\begin{equation} V_{\rm d}(z) =\frac{m}{2}\omega_z^2z_0^2+\frac{m}{2}\omega_z^2z^2-\frac{m}{2}\sqrt{4C^2+4\omega_z^4z^2z_0^2} \end{equation}
This potential quantitatively resembles the shape of individual harmonic wells around $\pm z_0$ up to terms proportional to $C^2$.

Specifically, for a given sufficiently high barrier this allows us to prepare zig-zag chains in each of the two wells 
which are the energetically lowest equilibrium configuration of the double well which we call the ground state configuration in the following
 Note that finding the many-particle minimum of a many ion system is generally a highly nontrivial task; hence the present choice of the 
double well potential is a crucial step to allow for a numerical study of the dynamics of coupled ion chains.
After preparing this configuration, i.e. its numerical determination, our strategy will be to ramp down the barrier height
by a certain amount which corresponds to a quench of the quantity $C$. Subsequently the resulting dynamics of the now
strongly coupled ion chains will be explored.
To understand the complex dynamics of coupled many-ion structures in non-equilibrium it is crucial to simplify our
model. First, since we are interested in the dynamics on large scales we neglect the micromotion. 
Second, we focus on a two-dimensional description, which simplifies the visualization of the ionic structures and their analysis but does not change 
qualitatively the resulting dynamics and phenomenology of the structure forming processes. 
Specifically, we choose parameters $\alpha=\omega_x/\omega_z \approx 8.3$ where the ground state configuration is a planar (2D) zig-zag structure in the $x-z$-plane far from the
transition to a helical (3D) zig-zag chain. The transition from the 2D structure to the 3D helical zig-zag chain occurs at $\alpha\approx 4.5$ ,
whereas the transition to a linear 1D line-structure occurs at $\alpha \approx 10$ (for $\omega_x=\omega_y$).  Thus, although ramping the barrier will generally produce both in-plane and out-of-plane fluctuations, the latter ones are generically small. 
A strong confinement in $y$-direction ($\omega_y/\omega_z\ge10$) prevents their amplification in the course of the consecutive dynamics; hence the 
ions stay close to the $x-z$ plane. Indeed the impact of small out-of plane fluctuations on the intra-plane dynamics 
is quadratically suppressed with the distance of the ions perpendicular to the $x-z$ plane 
\footnote {The only force which couples the dynamics of ions perpendicular to the $x-z$ plane to the intra-plane dynamics is the Coulomb coupling.
Consider two ions in distance $L=\sqrt{L_\parallel^2+L_\perp^2}$ where $L_\parallel$ is the projection of their distance onto the
$x-z$ plane and $L_\perp$ is their distance perpendicular to this plane. Then $F_\parallel = F \cos[\tan(L_\perp/L_\parallel)] \approx F[1-(L_\perp/L_\parallel)^2/2]$ 
for $L_\parallel\gg L_\perp$, showing that forces produced by small out-of plane oscillations onto the intra-plane dynamics are quadratically suppressed.} and consequently
a projection of the ion dynamics in 3D looks very similar to a direct 2D description. Hence, we focus on a two-dimensional minimal model given by the two-dimensional Hamiltonian:

\begin{align} H(\{ {\bf r}_i,{\bf p}_i\})=&\sum_{i=1}^n\frac{{{\bf p}_i}^2}{2m}+\sum_{i=1}^n \left[V_{d}(z)+V(x)\right]\nonumber \\
&+\sum_{i=1,j<i}^n \frac{Q^2}{4\pi\epsilon_0 r_{ij}}
\end{align}
with $V(x)=m\omega^2 x^2/2$, ${\bf r}_i=(x_i,z_i)$,\newline
$r_{ij}=\sqrt{(x_i-x_j)^2+(z_i-z_j)^2}$ and ${\bf p}=(p_x,p_z)$.
Introducing a rescaled time $t_u=1/\omega_z$ and space units $x_u=K\equiv[Q^2/(4\pi\epsilon_0m\omega_z^2)]^{1/3}$ and defining

\begin{align} 
&t^*=\omega_z t;\; x^*=\frac{x}{K};\; z^*=\frac{z}{K};\;z_0^*=\frac{z_0}{K};\;r_{ij}^*=\frac{r_{ij}}{K};\nonumber \\ 
&C^*=\frac{C}{K^2\omega_z^2};\; \alpha=\frac{\omega_x}{\omega_z}
\end{align}
we integrate the resulting Newtonian equations of motion. For the sake of convenience we will drop the stars in the following.
Finding the ground state configuration is a $2N$ dimensional optimization problem which is, for large $N$, numerically highly demanding, mainly 
because root finding algorithms converge, for most initial configurations, to some excited equilibrium configuration. 
To avoid this, we need a good initial guess for the positions of the ions which converges to the ground state configuration. 
Here, the quantitative similarity of both wells of the chosen double well with individual harmonic traps comes into play: it allows
us to exploit the existing knowledge on a single zig-zag ionic chain in the literature and hence to numerically efficiently
determine the double zig-zag ground state configuration within standard multidimensional rootfinding \cite{gsl}, which is visualized in Fig.~\ref{figa}. 
Note that the present setup also allows for a 'first excited' equilibrium configuration which is very similar to the ground state configuration (Fig.~\ref{figa}(b,c))
and whose energy exceeds that of the ground state {configuration by a factor of less than $10^{-10}$. We call this first excited equilibrium configuration the mirror configuration.

\section{Results and Analysis}
 We now use the planar double-zig-zag ground state configuration
as our initial configuration and assume that the initial velocities of all ions vanish which can be achieved approximately
via modern cooling techniques in experiments with trapped ions \cite{Drull1980,Lars1986,Died1989,Schl2006,Kell2006}.
We excite this initial configuration by performing a sudden quench of the barrier height between the two wells. This allows
in the course of the dynamics the zig-zag structures in both wells to move closer together, thereby effectively increasing their coupling.
Here we quench, exemplary, from $C=1.26\cdot 10^{-2}$ to $C=6.3$.
This quench leads to an energy excess of less than two percent with respect to the ground state configuration 
in the $C=6.3$ potential which is still a double zig-zag configuration. 
To explore the time evolution of the double zig-zag configuration after the quench we use an implicit Gaussian 4th order Runge-Kutta algorithm \cite{gsl} for 
integrating the Newtonian equations of motion.
For very short times we mainly observe a center of mass motion of both zig-zag structures towards each other, accompanied by small oscillations of the individual ions. 
At $t \sim 1.5$ the two chains begin to melt in the center of the system which continues until 
the zig-zag structures are completely destroyed.
Strikingly, as a result of the melting process, we do not only obtain phases of irregular oscillations of the ions in the trap (see for example Fig.~\ref{fig:Config} at $t=18.91$), but
also many phases of transient structures featuring an unexpected degree of order:
we observe the formation of ions in lines, arcs and cross like structures and partial revivals of
the dynamics (see Fig.~\ref{fig:Config}). As a result, coupled zig-zag chains not only exist as a stable ground and excited equilibrium configuration
but, once exposed to perturbations such as our quench, experience a complex reordering process with both irregular phases of motion and ordered configurations as quantified in Fig.~\ref{fig:circle} a).
It turns out that the shape of the specific configurations we observe is sensitive to the initial configuration:
Choosing the mirror configuration instead of the ground state configuration as our new initial configuration and performing the same quench of the barrier as before leads to a 
second set of ordered structures, including 
arc like configurations and elliptical arrangements of ions (see Fig.~\ref{fig:Config} b).

\begin{figure}[h!]
\begin{center}
\includegraphics[width=0.9\linewidth]{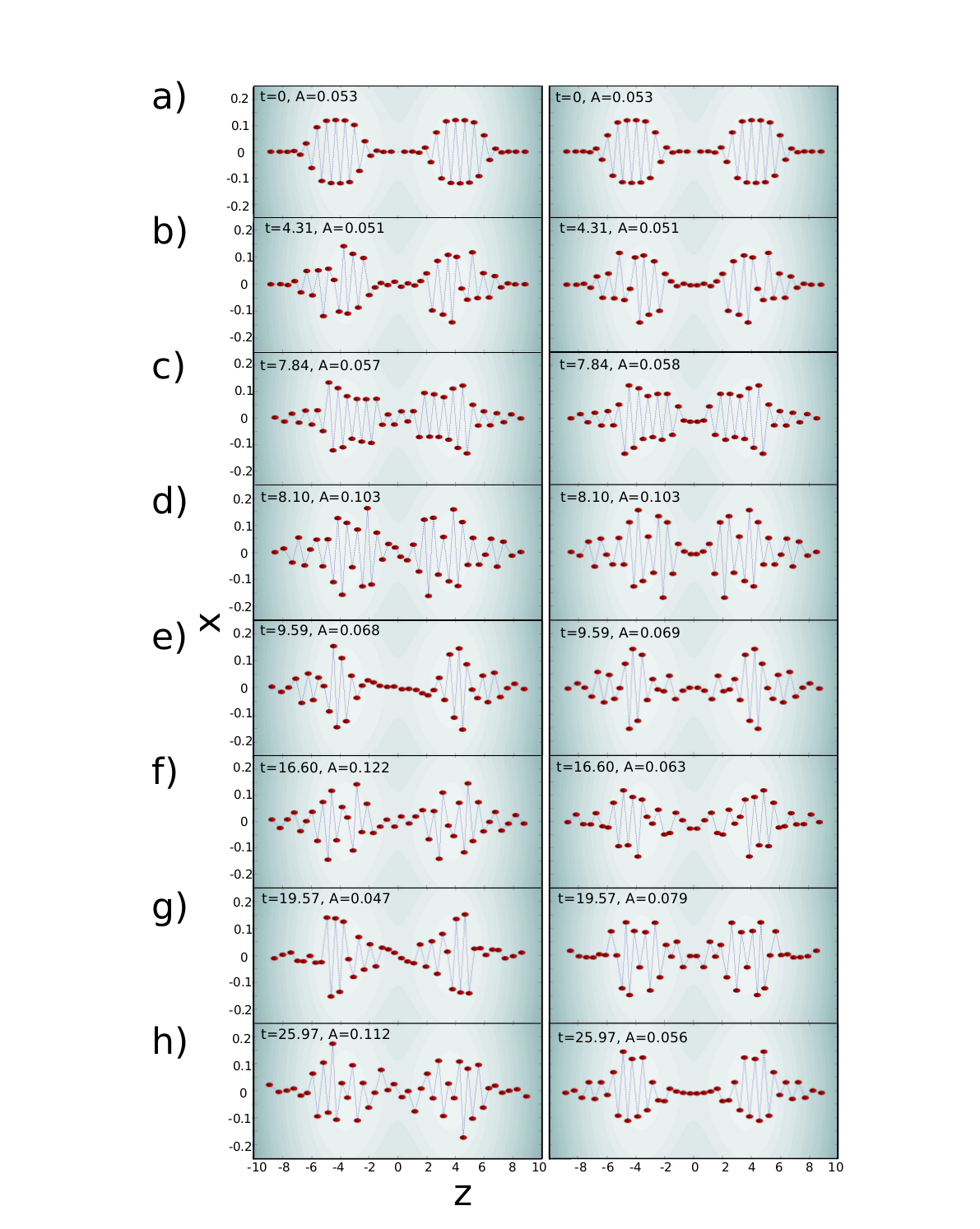} \\
\caption{Snapshots of the time evolution of $N=44$ ions in the double well for an initial configuration given by 
the ground state configuration (left column) and the mirror configuration (right column) before the quench.
Parameters (in scaled units): $\alpha=8.25; z_0=5/2^{1/3}$; $C$ is changed from $0.0126$ to $6.26$.}
\label{fig:Config}
\end{center}
\end{figure}
   
In the following we introduce a measure based on the idea of Voronoi diagrams \cite{Prep1985} to quantify the degree of order underlying the structures shown in Fig.~\ref{fig:Config}.
Specifically, when drawing a circle around each ion, whose  diameter is given by the distance to its nearest neighbor, our measure is 
defined as the sum of the areas of the respective circles around all $N=44$ ions, i.e. 
\begin{equation}
 A(t)=\frac{\pi}{4}\sum_i((x_i(t)-x_{j(t)}(t))^2+(z^*_i(t)-z^*_{j(t)}(t))^2).
\end{equation}

To evaluate this measure we rescaled the z-coordinates of all ions $z^*=z\cdot k$ with $k=0.04$, such that the average $x$-distance of adjacent ions 
equals their average distance in $z$-direction. The result of $A(t)$ is shown in Fig.~\ref{fig:circle} a) and features pronounced oscillations: all structures which 
appear to be ordered from a visual impression and show ions arranged in arcs, lines or crosses (Fig.~\ref{fig:Config})
lead to distinct minima in $A$, whereas 
irregular structures (Fig.~\ref{fig:Config} d (left and right), f (left) and g (right)) generate larger values of $A$.
Clearly A is a measure for the clustering of ions on one-dimensional manifolds such as straight lines, arcs or crosses  leading to low values of A,
whereas acquires larger values if the ions tend to cover a two dimensional area.\footnote{Note, that A would be also small if all ions would cluster, which is however excluded 
from the present situation due to the chain character of their arrangement.}
\begin{figure}[h!]
 \centering
 \includegraphics[bb=0 0 1179 854,scale=0.3,keepaspectratio=true]{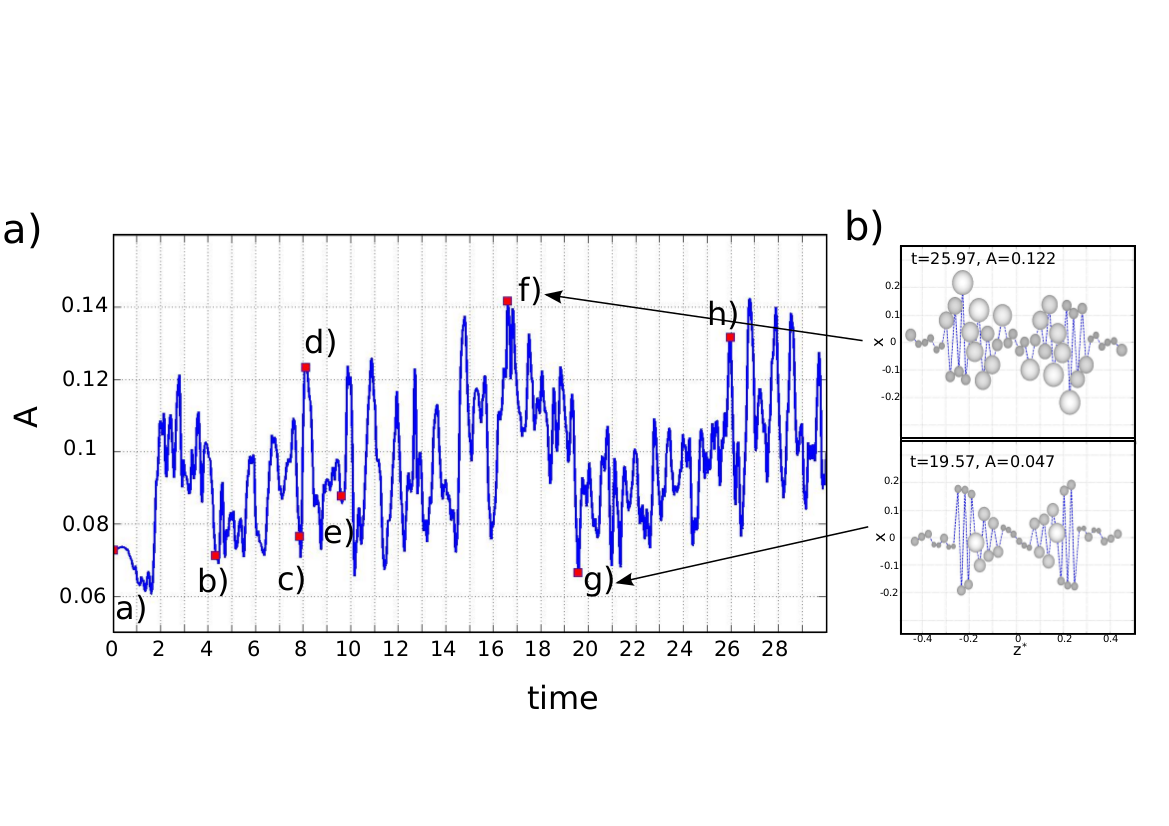}
 \caption{Time evolution of the measure A as defined in Eq. (5). Red squares show times at which snapshots in Fig. 2 were taken for the ground configuration(left column).
Insets illustrate the definition of A as the sum of the areas of all shown circles. Upper and lower insets illustrate how A accounts for
typical irregular and regular configurations respectively. }
 \label{fig:circle}
\end{figure}
   We now analyze the 
 observed emergence of transiently ordered ion structures originating from the melted double zig-zag structures.
Clearly, if the initial quench of the barrier had induced only small oscillations around the initial configuration of the ions 
 the dynamics could be described to a good approximation by the linearized equations of motion. In terms of normal modes \cite{Goldstein2014}, it would then consist of 
a superposition of harmonic oscillations with preserved
 oscillation amplitudes.
\\Here, the ion dynamics does not consist of small amplitude oscillations (the initial configuration dissolves completely) but is essentially nonlinear. Generally, nonlinearities couple 
different normal modes and transfer energy between them resulting in varying normal mode amplitudes.
Here, it turns out however, that a normal mode analysis is still useful: 
We find that during certain phases of the ion dynamics only a weak energy transfer between different normal modes takes place, indicating that nonlinearities
have only a weak impact on the ion dynamics during these phases. These phases are interrupted by short transition 'events', where significant population is transferred between 
different eigenvectors.
Specifically, we use standard techniques to numerically calculate the eigenfrequencies $\{\omega_i\}$
and the associated eigenvectors $\{{\bf E}_i \}$ around the new many-ion equilibrium configuration of the double well following the quench.
As the eigenvectors form a basis of the $N$-dimensional configuration space (here sorted by their eigenfrequencies)
we can now represent  position and velocity of each ion at a fixed time $t$ in this basis:

\begin{eqnarray} \left({\bf r}_1,..,{\bf r}_N\right)(t)&=&\left({\bf r}_1^{\rm eq},..,{\bf r}_N^{\rm eq}\right)+\sum_i p_i(t) {\bf E}_i; \nonumber 
 \\ \left({\bf v}_1,..,{\bf v}_N\right)(t)&=&
\sum_i k_i(t){\bf E}_i
\label{eqn:eigenv}
\end{eqnarray} 
Here, ${\bf r}^{\rm eq}_m$ is the position of the $m$-th ion in the ground state configuration (mirror configuration) of the double well following the quench
and $p_{i}^2(t)$ ($k_{i}^2(t)$) describe the population of the linear eigenvectors ${\bf E}_i$.
In the linear regime the population coefficients reduce to $p_i(t)=P_i \cos(\omega_i*t)$ and $k_i(t)=K_i \sin(\omega_i*t)$ with $P_i,K_i$ being constant amplitudes. 
\begin{figure}[h!]
\begin{center}
\includegraphics[width=0.8\linewidth]{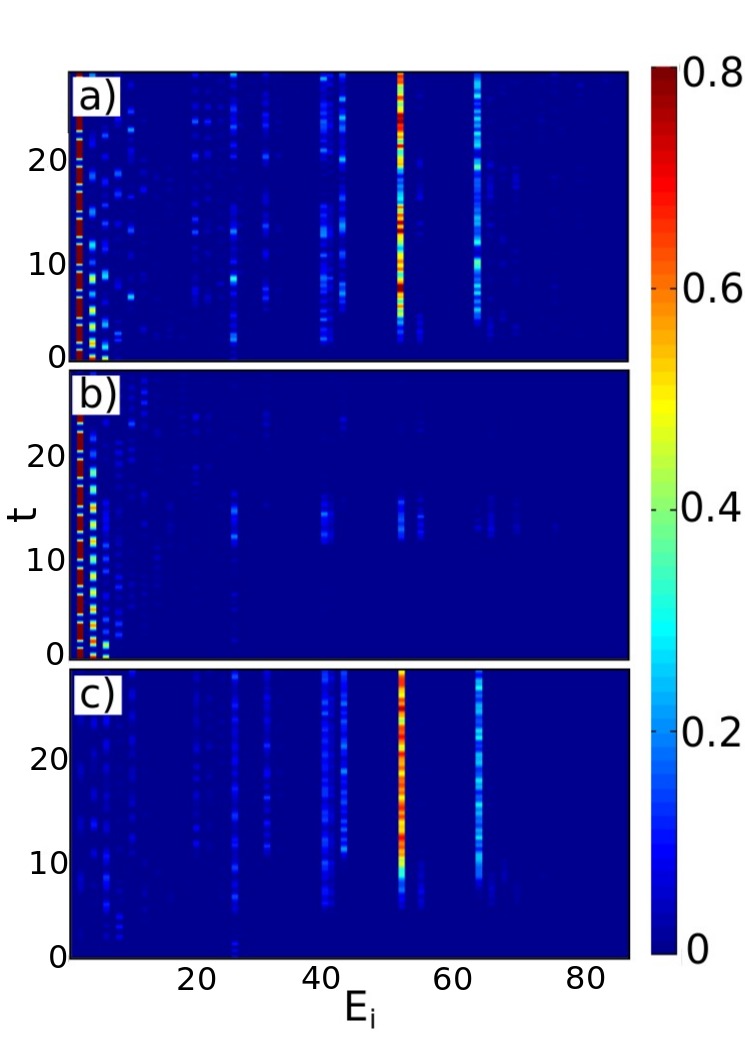}
\caption{Time-dependent population of linear eigenvectors defined as $p_i(t)=(d_i(t) \cos(\omega_i t))^2$ (color) for different initial populations:
a) created by the quench for the ground state configuration as initial configuration. 
b) as in a) but after setting all but the three largest initial populations ($p_2,p_4,p_6$) to zero.
c) Population of only one mode with $p_{26}(t=0)=0.11524$.
Parameters like in Fig.~\ref{fig:Config}.}
\label{fig:pcolor_1}
\end{center}
\end{figure}

\begin{figure}[h!]
\begin{center}
\includegraphics[width=0.8\linewidth]{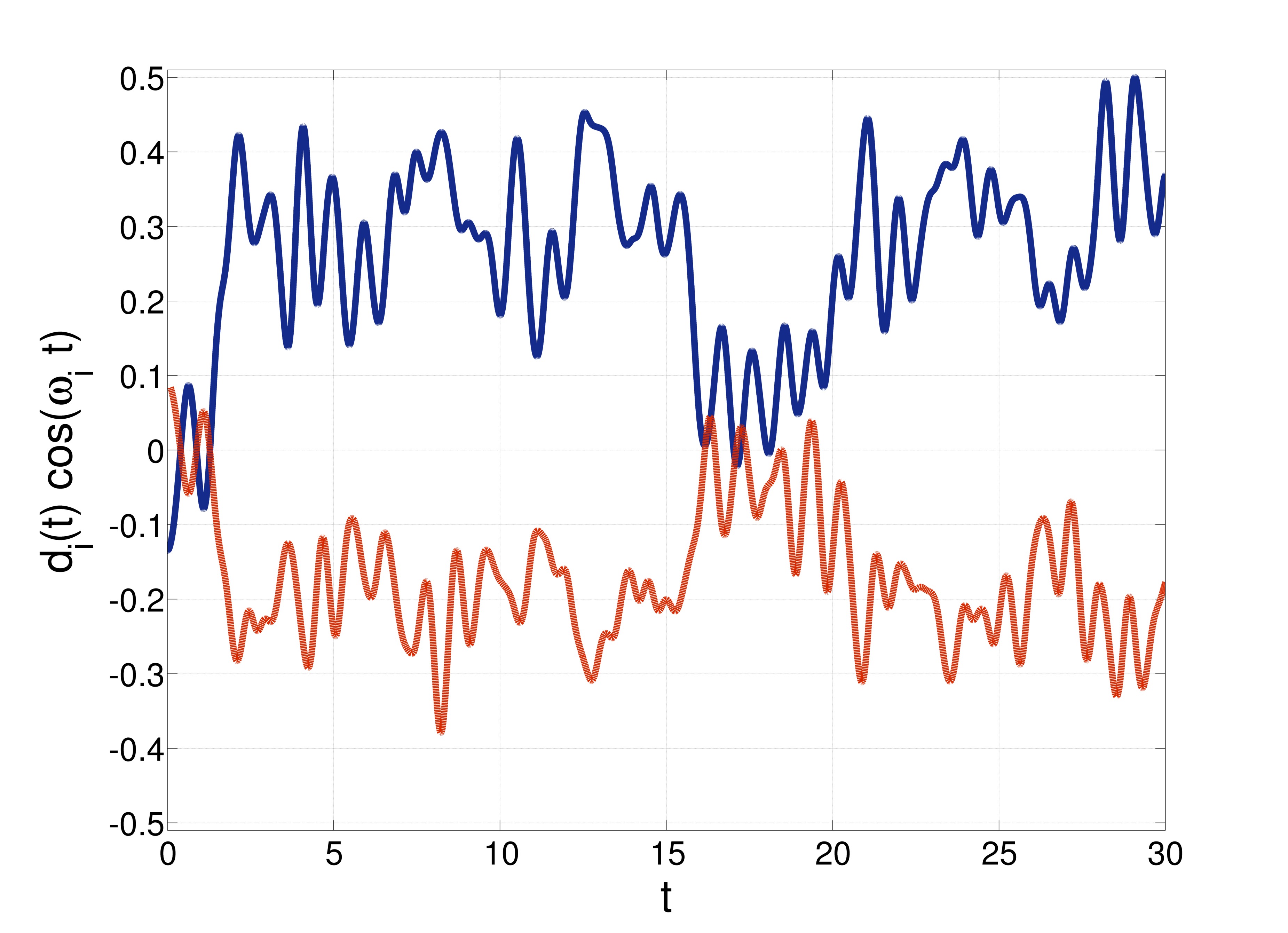}
\caption{Magnification of the population dynamics of $d_{40}(t) \cos(\omega_{40} t)$ and $d_{41}(t) \cos(\omega_{41} t)$ as defined in eq.\ref{eqn:eigenv}.}
\label{pairs}
\end{center}
\end{figure}
We now analyze the population dynamics of the eigenvectors ${\bf E}_i$  as induced by nonlinear effects. The initial quench mainly populates the 
vectors ${\bf E}_2,{\bf E}_4,{\bf E}_6$ (Fig.~\ref{fig:pcolor_1}a at $t=0$). 
These belong to the slowest modes whose population is allowed by symmetry (they possess the same symmetry as the initial configuration).
 Fig.~(\ref{fig:pcolor_1}a) shows that a first significant energy transfer to higher modes takes place at $t \sim 1.5$, which is the time where the zig-zag configurations start to melt, 
and then spreads to more and more other modes. 
Notably, the energy does not spread in an arbitrary manner over a whole band of modes which we would expect for a purely irregular or strongly chaotic system, but 
many of the eigenvectors either keep their population for relatively long times (Fig.~\ref{fig:pcolor_1}a) or exchange them only pairwise (Fig.~\ref{pairs}).
This indicates that the many ion dynamics is not irregular but possesses a high degree of order.
We now study in more detail how the population transfer between individual linear normal modes takes place. 
Therefore, we populate only one (or several) of the linear eigenvectors as our initial configuration and explore the resulting dynamics.
Populating for example only vector ${\bf E}_2$ which has the largest population after the quench, i.e. choosing
$\left({\bf r}_1,..,{\bf r}_N\right)(t=0)=\left({\bf r}_1^{\rm eq},..,{\bf r}_N^{\rm eq}\right)+d_2(t=0){\bf E}_2$
leads to a permanent 'breathing' oscillation of all ions in the left well and 
an opposite motion of the ions in the right well.
Accordingly, we observe almost no transfer of population to other eigenvectors and the considered
single mode excitation leads to persistent small and periodic oscillations rather than inducing a melting of the double zig-zag.
If we populate, e.g., the three eigenvectors which attract the strongest population by the quench of the potential barrier, we obtain a similar result:
Only very few energy is transferred to other modes (Fig.\ref{fig:pcolor_1}b).
There are however eigenvectors belonging to higher modes, whose initial population, even if it is small, induce a hierarchical spreading of population between different eigenvectors. 
In particular, if we weakly populate vector
${\bf E}_{26}$ above $p_{26}(t=0)$ we obtain a population dynamics which reproduces
large parts of the occupation dynamics as induced by the quench of the potential in the full problem (Fig.~\ref{fig:pcolor_1}c).
Already a weak population of this vector does therefore lead to significant nonlinear effects 
strongly mixing the linear eigenmodes. As a result, it depends on the details of how we excite the double zig-zag ground state configuration whether it
responds with small oscillations or with a complete melting and the dynamical reconfiguration of the ions into ordered
(transient) structures.
 Here, we neglected micromotion for simplicity. Accounting for it would however influence the occupation of eigenvectors \cite{Kauf2012}, but
presumably only on relatively long time scales. Accordingly, it could be interesting to explore if micromotion 
leads to a diffusion-like spreading of population over many eigenvectors and therefore 
generates a slow crossover from the partly regular dynamics quantified above to an increasingly irregular ion dynamics on long timescales.
\\Let us finally address the experimental realization of our setup employing state of the art ion technology.
Typical experimental parameters for segmented Paul traps are $\omega_{rf}/2\pi=4.2-50 \text{MHz}$, $U_{rf}=8-350\text{V}$ with applied dc voltages in axial direction of up to $10\text{V}$ \cite{Kauf2014,Hens2006}.
Depending on the ion species and trap design this results in radial direction in $\omega/2\pi=1-5\text{MHz}$ and in axial direction in $\omega_z/2\pi=0-5\text{MHz}$.
For the dynamics only the frequency ratio $\alpha=\frac{\omega}{\omega_z}=8.25$ is of importance (e.g. $\omega/2\pi=4.5\text{MHz}$
and $\omega_z/2\pi=0.545\text{MHz}$). 
The parameters $z_0$ and $C$ depend on the dc voltage and the trap geometry and are in the range of $\unit{30}{\micro \meter} $ for the well position $z_0$ and up to 
$\unit{300}{\micro \meter}^2 \cdot \text{MHz}^2$ for $C$.
The ion configurations in the course of the dynamics could be detected by fluorescence light with a CCD camera \cite{Miel2012,Kauf2014}.

\section{Conclusions}
While structure formation with trapped ions is already a versatile and active research area, 
recent experimental progress suggests the coupling of individual ionic structures in multi-segmented traps.
The present work elaborates, on a minimal model that allows for coupled zig-zag ion chains that respond in 
a highly selective way when we excite them: Depending on the details of the excitation we either obtain regular oscillations around the equilibrium configuration 
or a complete melting followed by a sequence of structures with ions arranged in lines, arcs and cross like formations. 
Our model can be extended to explore the dynamics of whole arrays of coupled ionic structures in multisegmented traps.
The selective response of coupled zig-zag chains to weak excitations promises an energy transport (or sound propagation) along the trap which strongly depends on  
the initial signal. Particularly in cases where the ions respond with small oscillations we expect a much weaker energy transport than in cases of
complete melting where nonlinear effects couple linear eigenmodes and convert Coulomb energy to intersite 
from one site of the double well to the other one motion of ions. 
A further application of coupled ionic structures in multiwell traps could 
exploit the current interest in topological defects (kinks) in zig-zag structures to study their collision 
and transferability between different individual zig-zag chains. 

\section*{Acknowledgments}
We thank Christoph Petri, Sven Kroenke, Jan Stockhofe, Fotis K. Diakonos for useful discussions and suggestions. A.K. thanks Stephan Klumpp for scientific discussions and support. 
B.L. gratefully acknowledges funding by a Marie Sk\l{}odowska-Curie Intra European Fellowship (G.A. no. 654908) within Horizon 2020. 

\bibliography{AKtrappedIons}

\begin{thebibliography}{10}

\bibitem{Mori20041}
{\em Physical Review Letters}, 93, 2004.

\bibitem{Balt20121}
J.D. Baltrusch, C.~Cormick, and G.~Morigi.
\newblock {\em Physical Review A}, 86:032104, 2012.

\bibitem{Boll1994}
J.~J. Bollinger, D.~J. Wineland, and D.~H.~E. Dubin.
\newblock {\em Physics of Plasmas}, 1:1403, 1994.

\bibitem{Boni2008}
M.~Bonitz, P.~Ludwig, H.~Baumgartner, C.~Henning, A.~Filinov, D.~Block, O.~Arp,
  A.~Piel, S.~K{{\"a}}ding, Y.~Ivanov, A.~Melzer, H.~Fehske, and V.~Filinov.
\newblock {\em Phys. Plasmas}, 15:055704, 2008.

\bibitem{Died1989}
F.~Diedrich, J.C. Bergquist, Wayne~M. Itano, and D.J. Wineland.
\newblock {\em Phys. Rev. Lett.}, 62:403, 1989.

\bibitem{Died1987}
F.~Diedrich, E.~Peik, J.M. Chen, W.~Quint, and H.~Walther.
\newblock {\em Physical Review Letters}, 59:2931, 1987.

\bibitem{Drew1998}
M.~Drewsen, C.~Brodersen, L.~Hornekaer, J.S. Hangst, and J.P. Schiffer.
\newblock {\em Physical Review Letters}, 81:2878, 1998.

\bibitem{Drull1980}
R.E. Drullinger, D.J. Wineland, and J.C. Bergquist.
\newblock {\em Applied Physics}, 22:365, 1980.

\bibitem{Dub1999}
Daniel~H.E. Dubin and T.M. O'Neil.
\newblock {\em Reviews of Modern Physics}, 71:8, 1999.

\bibitem{Dub1993}
D.H.E. Dubin.
\newblock {\em Physical Review Letters}, 71:2753, 1993.

\bibitem{gsl}
M.~Galassi, J.~Davis, J.~Theiler, B.~Gough, G.~Jungman, P.~Alken, M.~Booth, and
  F.~Rossi.
\newblock {\em GNU Scientific Library Reference Manual}.
\newblock Network Theory Ltd., 3rd edition, 2009.

\bibitem{Ger1992}
D.~Gerlich.
\newblock {\em Advances in Chemical Physics Series}, 82:1, 1992.

\bibitem{Goldstein2014}
H.~Goldstein, C.P. Poole, and J.L. Safko.
\newblock {\em Classical Mechanics: Pearson New International Edition}.
\newblock Harlow, 2014.

\bibitem{Gui2015}
N.D. Guise, S.D. Fallek, K.E. Stevens, K.R. Brown, C.~Volin, A.W. Harter, J.M.
  Amini, R.E. Higashi, S.T. Lu, H.M. Chanhvongsak, T.A. Nguyen, M.S. Marcus,
  T.R. Ohnstein, and D.W. Youngner.
\newblock {\em Appl. Phys.}, 17:174901, 2015.

\bibitem{Blatt2008}
H.~H{{\"a}}ffner, C.~F. Roos, and R.~Blatt.
\newblock {\em Physics Reports}, 469:155, 2008.

\bibitem{Hens2006}
W.K. Hensinger, S.~Olmschenk, D.~Stick, D.~Hucul, M.~Yeo, M.~Acton,
  L.~Deslauriers, C.~Monroe, and J.~Rabchuk.
\newblock {\em Applied Physics Letters}, 88:034101, 2006.

\bibitem{Drew2001}
L.~Hornekaer, N.~Kjaergaard, A.M. Thommesen, and M.~Drewsen.
\newblock {\em Physical Review Letters}, 86:1994, 2001.

\bibitem{Hugh2011}
M.D. Hughes, B.~Lekitsch, J.A. Broersma, and W.K. Hensinger.
\newblock {\em Contemporary Physics}, 52:505, 2011.

\bibitem{Kauf2014}
H.~Kaufmann, T.~Ruster, C.T. Schmiegelow, F.~Schmidt-Kaler, and U.G.
  Poschinger.
\newblock {\em New Journal of Physics}, 16:073012, 2014.

\bibitem{Kauf2012}
H.~Kaufmann, S.Ulm, G.Jacob, U.~Poschinger, H.~Landa, A.~Retzker, M.B. Plenio,
  and F.~Schmidt-Kaler.
\newblock {\em Physical Review Letters}, 109:263003, 2012.

\bibitem{Kell2006}
A.~Kellerbauer and J.~Walz.
\newblock {\em New Journal of Physics}, 8:45, 2006.

\bibitem{Drew2003}
N.~Kjaergaard and M.~Drewsen.
\newblock {\em Physical Review Letters}, 91:095002, 2003.

\bibitem{Lars1986}
D.J. Larson, J.C.Bergquist, J.J. Bollinger, W.M. Itano, and D.J. Wineland.
\newblock {\em Physical Review Letters}, 57:70, 1986.

\bibitem{Lee2011}
T.E. Lee and M.C. Cross.
\newblock {\em Physical Review Letters}, 106:143001, 2011.

\bibitem{Liebchen2012}
B.~Liebchen, F.K. Diakonos, and P.~Schmelcher.
\newblock {\em New J. Phys.}, 14:103032, 2012.

\bibitem{Liebchen2014}
B.~Liebchen and P.~Schmelcher.
\newblock {\em Physical Review Letters}, 112:134102, 2014.

\bibitem{Liebchen2015}
B.~Liebchen and P.~Schmelcher.
\newblock {\em New Journal of Physics}, 17:083011, 2015.

\bibitem{Mad2004}
M.J. Madsen, W.K. Hensinger, D.~Stick, J.A. Rabchu, and C.~Monroe.
\newblock {\em Applied Physics B}, 78:639, 2004.

\bibitem{Miel2012}
M.~Mielenz, J.~Brox, S.~Kahra, G.~Leschhorn, M.~Albert, T.~Schaetz, H.~Landa,
  and B.~Reznik.
\newblock {\em Physical Review Letters}, 110:133004, 2013.

\bibitem{Mori20042}
G.~Morigi and S.~Fishman.
\newblock {\em Physical Review E}, 70:066141, 2004.

\bibitem{Prep1985}
F.P. Preparata and M.I. Shamo.
\newblock {\em Computational Geometry}.
\newblock Springer Verlag, New York, 1985.

\bibitem{Rust2014}
T.~Ruster, C.~Warschburger, H.Kaufmann, C.T. Schmiegelow, A.~Walther,
  M.~Hettrich, A.~Pfister, V.~Kaushal, F.~Schmidt-Kaler, and U.G. Poschinger.
\newblock {\em Physical Review A}, 90:033410, 2014.

\bibitem{Schiff1993}
J.P. Schiffer.
\newblock {\em Physical Review Letters}, 70:818, 1993.

\bibitem{Schl2006}
S.~Schlemmer, T.~Kuhn, E.~Lescop, and D.~Gerlich.
\newblock {\em International Journal of Mass Spectrometry}, 185:589, 1999.

\bibitem{Scha2010}
C.~Schneider, M.~Enderlein, T.~Huber, and T.~Schaetz.
\newblock {\em Nature Photonics}, 4:772, 2010.

\bibitem{Schu2006}
S.~Schulz, U.G. Poschinger, K.~Singer, and F.~Schmidt-Kaler.
\newblock {\em Fortschritte in der Physik}, 54:648, 2006.

\bibitem{Schu2008}
S.A. Schulz, U.G. Poschinger, F.~Ziesel, and F.~Schmidt-Kaler.
\newblock {\em New Journal of Physics}, 10:045007, 2008.

\bibitem{Stre2006}
A.I. Streltsov, O.E. Alon, and L.S. Cederbaum.
\newblock {\em Phys. Rev. A}, 73:063626, 2006.

\bibitem{Tana2014}
U.~Tanaka, K.~Suzuki, Y.~Ibaraki, and S.~Urabe.
\newblock {\em Journal of Physics B}, 47:035301, 2014.

\bibitem{Wilp2012}
G.~Wilpers, P.~See, P.~Gill, and A.G. Sinclair.
\newblock {\em Nature Nanotechnology}, 7:572, 2012.

\bibitem{Wine1987}
D.J. Wineland, J.C. Bergquist, W.~M. Itano, J.J. Bollinger, and C.H. Manney.
\newblock {\em Physical Review Letters}, 59:2935, 1987.

\end{thebibliography}

\end{document}